\documentclass[pra,showpacs,twocolumn]{revtex4}
\usepackage{graphicx,float,calc}
\usepackage{amsmath,amsfonts,amssymb}
\usepackage{color}
\usepackage{bm}
\usepackage[colorlinks,urlcolor=blue,citecolor=blue,linkcolor=blue]{hyperref}

\begin{document}
\title{Thermodynamic Properties of Rashba Spin-Orbit-Coupled Fermi Gas}
\author{Zhen Zheng$^{1,2}$}
\author{Han Pu$^{3,4}$}
\thanks{hpu@rice.edu}
\author{Xubo Zou$^{1,2}$}
\thanks{xbz@ustc.edu.cn}
\author{Guangcan Guo$^{1,2}$}
\affiliation{$^1$Key Laboratory of Quantum Information, University of Science and Technology of China, Hefei, Anhui, 230026, China}
\affiliation{$^2$Synergetic Innovation Center of Quantum Information and Quantum Physics, University of Science and Technology of China, Hefei, Anhui 230026, China}
\affiliation{$^3$Department of Physics and Astronomy and Rice Quantum Institute, Rice University, Houston, TX 77251-1892, USA}
\affiliation{$^4$Center for Cold Atom Physics, Chinese Academy of Sciences, Wuhan 430071, China}

\begin{abstract}
We investigate the thermodynamic properties of a superfluid Fermi gas subject to Rashba spin-orbit coupling and effective Zeeman field. We adopt a T-matrix scheme that takes beyond-mean-field effects, which are important for strongly interacting systems, into account. We focus on the calculation of two important quantities: the superfluid transition temperature and the isothermal compressibility. Our calculation shows very distinct influences of the out-of-plane and the in-plane Zeeman fields on the Fermi gas. We also confirm that the in-plane Zeeman field induces a Fulde-Ferrell superfluid below the critical temperature and an exotic finite-momentum pseudo-gap phase above the critical temperature.
\end{abstract}
\maketitle

\section{INTRODUCTION}

Ultracold Fermi gas \cite{giorgini_rmp2008},
with tunable atom-atom interaction through Feshbach resonance \cite{chin_rmp2010}, has been an ideal platform for the study of the crossover physics from weak-coupling Bardeen-Cooper-Schrieffer (BCS) pairing to a Bose-Einstein condensate (BEC) of bound pairs \cite{regal_prl2004,zwierlein_prl2004,bartenstein_prl2004}.
Recently synthetic gauge fields \cite{lin_nat2009} and spin-orbit (SO) coupling \cite{lin_nat2011, ji_nphys2014, chunlei_pra2013, pengjun_prl2012, cheuk_prl2012} were realized in experiments,
opening up a completely new avenue of research in superfluid Fermi gases.
The interplay of Zeeman fields and SO coupling leads to many novel phenomena at both zero
\cite{gong_prl2011,han_pra2012,seo_pra2012,jiang_pra2011,zengqiang_prl2011,zhoupra2013}
and finite temperatures \cite{lindong_njp2013,hu_njp2013},
from mixed singlet-triplet pairing to topological phase transition
\cite{gong_prl2011,xiaji_pra2012,wu_prl2013,zheng_pra2013,chunlei_natc2013,weizhang_natc2013}.

Many of the interesting physics of SO-coupled Fermi gas can be captured by mean-field theory. However, it is also true that mean-field theory may fail under many circumstances. For example, mean-field theory, which does not include the effect of noncondensed pairs,
fails to describe accurately the phase transition from a superfluid to a normal gas, particularly for systems with strong interaction.
Beyond-mean-field theoretical methods have been proposed  \cite{nozieres1985,melo_prl1993,ohashi_prl2002,machida_pra2006,xiaji_pra2005}.
Here we present a theoretical investigation under the framework of the T-matrix scheme to address the superfluid properties of a Rashba SO coupled Fermi gas over the entire BCS-BEC crossover regime \cite{qijin_prl1998,maly1999,loktev_pr2001,perali_prb2007,qijin_pr2005,kinnunen,huihu_prl2010,qijin,kinast, stajic,bauer_prl2014, haoguo_arxiv2013,lianyi_pra2013,zhangjing}.
In the absence of the Zeeman field, it was shown that the SO coupling enhances superfluid pairing \cite{renyuan_prl2012,zengqiang_prl2011,lianyi_prl2012,lianyi_pra2013}. At the mean-field level, we know that
the presence of both the SO coupling and a perpendicular (out-of-plane) Zeeman field give rise to effective $p$-wave pairing \cite{tewari_prl2007,chuanwei_prl2012}.
Meanwhile introducing an in-plane Zeeman component creates an anisotropic Fermi surface which favors finite-momentum pairing,
giving rise to a Fulde-Ferrell superfluid \cite{wu_prl2013,zheng_pra2013,chunlei_natc2013,weizhang_natc2013,ff1,ff4}.
These previous studies motivated our present work, in which we investigate the thermodynamic properties of an SO coupled Fermi gas subject to a Zeeman field. We will focus our calculation on two important quantities that are measurable in experiment: the superfluid to normal transition temperature and the isothermal compressibility.

We organize the paper as follows.
In Sec.~\ref{sec_model_hamiltonian}, we give an introduction to the Rashba SO coupled model
and briefly describe the T-matrix scheme used in the calculation.
Then in Sec.~\ref{sec_zero-temperature_properties},
we briefly review the zero-temperature properties of the system.
The superfluid transition temperature $T_c$ in the BEC-BCS crossover is investigated in Sec.~\ref{sec_superfluid_transition_temperature}, and its
dependence on the Zeeman field is emphasized. We present the numerical results of compressibility in Sec.~\ref{sec_isothermal_compressibility} before we provide a summary in Sec~\ref{summary}. We show that both the superfluid transition temperature and the compressibility have distinct dependence on the out-of-plane and the in-plane Zeeman fields.
In the Appendix we present technical details of the T-matrix formalism.

\section{Model}
\label{sec_model_hamiltonian}

We consider a three-dimensional two-component degenerate Fermi gas with Rashba SO coupling together with effective Zeeman fields.
This system can be described by the following Hamiltonian:
\begin{eqnarray}
\mathcal{H} &=& \int d{\bf r}\,
\psi^\dag ( \mathcal{H}_0 + \mathcal{H}_\mathrm{so} + h_z\sigma_z +h_x\sigma_x) \psi({\bf r}) \nonumber\\
&& + U \int d{\bf r}\, \psi^\dag_\uparrow({\bf r}) \psi^\dag_\downarrow({\bf r})
\psi_\downarrow({\bf r}) \psi_\uparrow({\bf r})~,
\end{eqnarray}
where $\psi({\bf r})=(\psi_\uparrow \,,\psi_\downarrow)^T$ represents the fermionic field operator and $\mathcal{H}_0 = -\nabla^2/(2m)-\mu$ represents the kinetic energy
with $\mu$ being the chemical potential.
The Rashba SO-coupling term takes the form
$\mathcal{H}_\mathrm{so} = \alpha ( k_y\sigma_x - k_x\sigma_y )$ in the $xy$-plane, with the parameter $\alpha$ characterizing the strength of the SO coupling.
We consider both an out-of-plane Zeeman field $h_z$ and an in-plane Zeeman field $h_x$.
The quantity $U$ represents the bare two-body interaction constant and in the calculation will be replaced by the $s$-wave scattering length $a_s$ through the standard regularization scheme:
$1/U = m/(4\pi\hbar^2a_s)-\sum_k m/k^2$.

In the mean-field BCS theory, we introduce an order parameter
$\Delta=U\sum_{\bm{k}}\langle \psi_{\bm{Q}+\bm{k},\uparrow}\psi_{\bm{Q}-\bm{k},\downarrow}\rangle$ to characterize the property of the superfluid. Here $\bm{Q}$ represents the center-of-mass momentum of the pairs. A finite ${\bm{Q}}$ arises from the presence of the in-plane Zeeman field \cite{wu_prl2013,zheng_pra2013,chunlei_natc2013,ff1,lindong_njp2013,hu_njp2013,ff4}.
At finite temperature in a T-matrix scheme, the order parameter $\Delta(T)$ can be divided into two parts:
$\Delta^2 = \Delta_\mathrm{sc}^2 + \Delta_\mathrm{pg}^2$ \cite{lianyi_pra2013}. Here $\Delta_{\rm sc}$ is the superfluid gap arising from the condensed pairs and vanishes above the superfluid transition temperature $T_c$.
The pseudo-gap $\Delta^2_\mathrm{pg}\sim\langle\Delta^2(T)\rangle - \langle\Delta(T)\rangle^2$ describes the thermodynamic fluctuation of non-condensed pairs \cite{qijin_prl1998,kosztin_prb2000}.
Below the superfluid transition temperature $T_c$, the thermodynamic quantities are determined
by the Thouless criterion \cite{thouless_1960} within the T-matrix scheme instead of the BCS formalism in the mean-field model.
This is because at finite temperature thermodynamic fluctuation plays a critical role with a tendency towards destroying the pairing condensation.
The Thouless criterion for finite pairing momentum $\bm{Q}$ takes the form:
\begin{equation}
U^{-1}+\chi(0,\bm{Q})=0 ~,
\end{equation}
where $\chi(0,\bm{Q})$ is the spin symmetrized pair susceptibility.
More technical details are given in the Appendix.

The T-matrix scheme adopted here was first developed in the context of high-$T_c$ cuprates \cite{qijin_prl1998,maly1999,loktev_pr2001,perali_prb2007}. It was later applied to study the BEC-BCS crossover phenomenon in ultracold atomic Fermi gases \cite{qijin_pr2005}. The T-matrix theory was found to be reasonably successful in providing theoretical support for the measured radio-frequency (rf) spectrum \cite{kinnunen,huihu_prl2010,qijin}, density profiles \cite{stajic}, and thermodynamic properties such as heat capacity \cite{kinast}, pressure \cite{bauer_prl2014}, and isothermal compressibility \cite{haoguo_arxiv2013} of ultracold Fermi gases. More recently, this theory was also applied to Fermi gases with spin-orbit coupling \cite{lianyi_pra2013, zhangjing}. In Ref.~\cite{zhangjing}, it was found that the theoretical calculation qualitatively agrees with the measured rf spectrum of a spin-orbit-coupled Fermi gas on the BEC side of the resonance. Given these past successes, we are confident that the T-matrix scheme indeed represents a vast improvement over the simple mean-field theory and should be at least qualitatively valid over the whole BEC-BCS crossover regime.

\begin{figure}[htbp]
\centering
\includegraphics[width=0.35\textwidth]{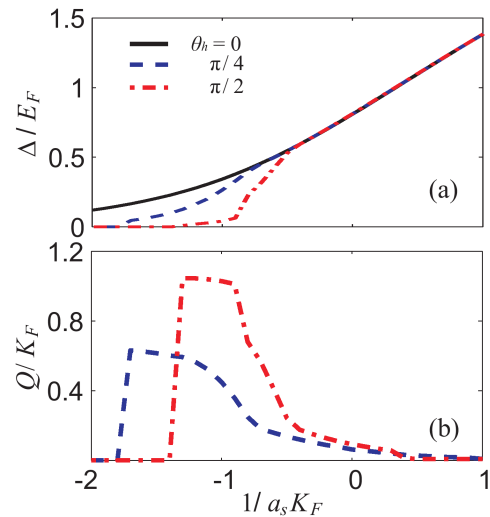}
\caption{(Color online) Thermodynamic quantities at zero temperature,
(a) order parameter, and (b) pairing momentum, as functions of interaction strength characterized by $1/a_s K_F$ for different $\theta_h$. Here the SO coupling strength is
$\alpha K_F=2.0E_F$, and the effective Zeeman field strength is $h=0.5E_F$. $\theta_h$ is the angle between the effective Zeeman field and the $z$-axis, such that $h_z = h \cos \theta_h$ and $h_x = h \sin \theta_h$. Therefore $\theta_h=0$ represents a pure out-of-plane Zeeman field, while $\theta_h=\pi/2$ represents a pure in-plane Zeeman field. $K_F$ and $E_F$ are the Fermi wave number and Fermi energy of the ideal Fermi gas, respectively.}
\label{crossover_para}
\end{figure}

\section{Zero-temperature properties}
\label{sec_zero-temperature_properties}

We first briefly review the main features of the system at zero temperature. Note that the pseudo-gap tends to zero in the low-temperature limit as $\Delta_\mathrm{pg}\sim T^{3/4}$ [see Eq.~(\ref{delta_pg})] and vanishes at $T=0$. Hence, the T-matrix scheme we adopted here reduces to the mean-field theory at zero temperature. The Rashba SO coupling and the Zeeman field have opposite effects on the magnitude of the gap parameter: the former tends to enhance the gap, while the latter tends to reduce the gap. Fig.~\ref{crossover_para}(a) displays the gap parameter as a function of interaction strength for several different Zeeman fields. The in-plane Zeeman field is known to break the symmetry of the band structure and results in Cooper pairs with finite momentum, a signature of the Fulde-Ferrell superfluid. In Fig.~\ref{crossover_para}(b), we show how the magnitude of the momentum of Cooper pairs $Q=|\bm{Q}|$ ($\bm{Q}$ is along the $y$-axis \cite{zheng_pra2013}) changes in the BEC-BCS crossover. $Q$ decreases quickly on the BEC side of the resonance as the two-body $s$-wave interaction, which favors zero-momentum pairing, dominates in that regime.

\begin{figure}[htbp]
\centering
\includegraphics[width=0.35\textwidth]{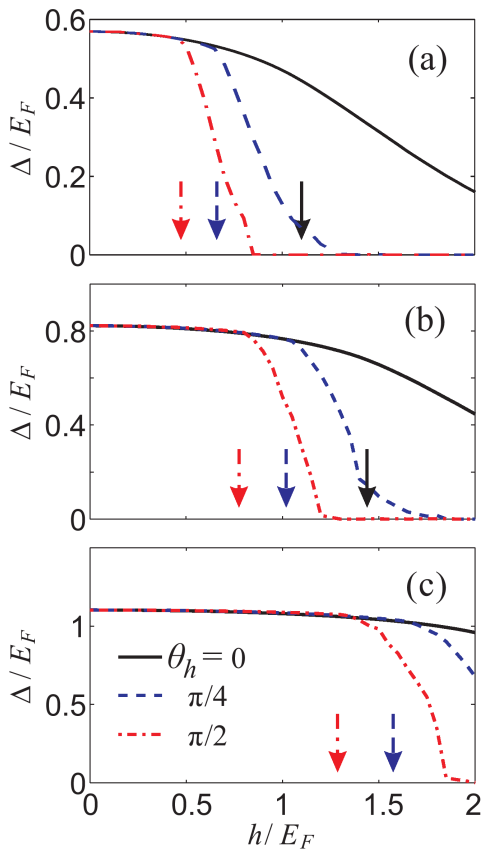}
\caption{(Color online) Superfluid gap at $T=0$ as a function of the Zeeman field strength for (a) $1/a_sK_F= -0.5$, (b) 0, and (c) 0.5. The solid, dashed, and dot-dashed lines correspond to $\theta_h= 0$, $\pi/4$, and $\pi/2$, respectively. The SO-coupling strength is
$\alpha K_F=2.0E_F$. The vertical arrows indicate the critical Zeeman field strength at which the system becomes gapless.}
\label{deltah}
\end{figure}

In Fig.~\ref{deltah}, we show how the zero-temperature superfluid gap varies as the Zeeman field strength changes. For a weak Zeeman field, the gap is insensitive to the orientation of the field. In general, $\Delta$ decreases as $h=\sqrt{h_z^2+h_x^2}$ increases due to the pair-breaking effect of the Zeeman field. However, as $h$ exceeds some threshold value, the decrease of the gap becomes sensitive to the orientation of the field: The larger the in-plane Zeeman field component is, the steeper the decrease of the gap becomes. As we will show below, this threshold value corresponds to the field strength at which the quasi-particle excitation gap vanishes.

\begin{figure}[htbp]
\centering
\includegraphics[width=0.32\textwidth]{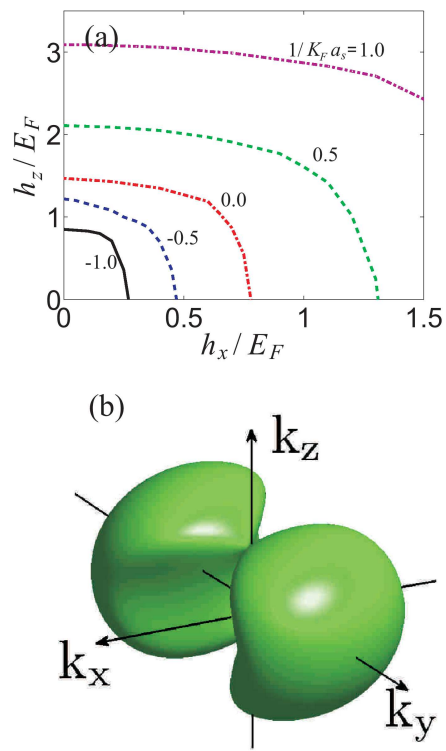}
\caption{(Color online) (a) The critical magnetic field that separates the gapped region and the gapless region. The region below the lines is gapped. Here the SO-coupling strength is
$\alpha K_F=2.0E_F$. (b) The nodal surface in the gapless region for $h_z=0.0$, $h_x=1.0E_F$, $\alpha K_F=2.0E_F$, and $1/a_sK_F=0.0$. The range of the plot is from $-2 K_F$ to $2K_F$ along each direction.}
\label{phase}
\end{figure}

Another important feature induced by the Zeeman field is that it closes the bulk quasi-particle excitation gap at a critical value. Fig.~\ref{phase}(a) represents a phase diagram in which we plot the critical Zeeman field at which the system changes from gapped to gapless. The region below each line represents the gapped phase. As the interaction strength varies from the BCS side to the BEC side, the order parameter increases [see Fig.~\ref{crossover_para}(a)], and correspondingly the critical field strength increases and the gapped region enlarges. In the absence of the in-plane Zeeman field (i.e., when $h_x=0$), the system becomes gapless due to the presence of the discrete Fermi points \cite{gong_prl2011} located along the $k_z$-axis in momentum space at $k_z=\pm\sqrt{2m(\mu\pm\sqrt{h^2-\Delta^2})}$. These Fermi points are topological defects. Hence the transition from the gapped to the gapless region in this case also represents a transition from a topologically trivial to a topologically nontrivial quantum phase. For finite $h_x$, by contrast, the gapless region features a nodal surface in momentum space on which the single particle excitation gap vanishes \cite{lindong_njp2013,zhoupra2013,yong_prl2014}. An example of such a nodal surface is plotted in Fig.~\ref{phase}(b). The vertical arrows in Fig.~\ref{deltah} indicate the critical Zeeman field strength at which the system becomes gapless. One can see that the sharp drop of the superfluid gap is correlated with the appearance of the nodal surface induced by a large in-plane Zeeman field. As we shall show below, the presence of the nodal surface also has dramatic effects on the thermodynamic properties of the system.

\begin{figure}[htbp]
\centering
\includegraphics[width=0.48\textwidth]{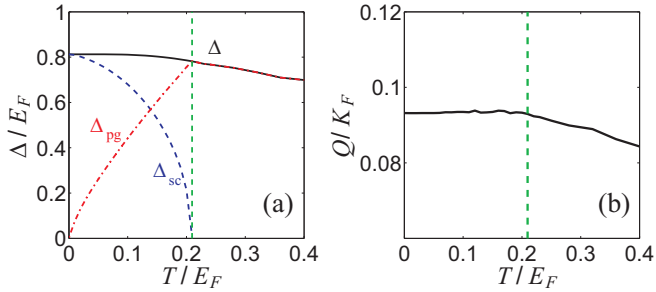}
\caption{(Color online) (a) The total gap $\Delta$, pseudo-gap $\Delta_{\rm pg}$ and superfluid gap $\Delta_{\rm sc}$ as functions of the temperature.
(b) The corresponding pairing momentum $Q$ as a function of the temperature.
The parameters used are
$1/a_sK_F=0.0$, $\alpha K_F=2.0E_F$, $h_z=0.0$, and $h_x=0.5E_F$. In this example, the green vertical dashed line indicates the critical temperature $T_c=0.210E_F$.}
\label{delta}
\end{figure}

\section{Superfluid Transition Temperature}
\label{sec_superfluid_transition_temperature}

We now turn our attention to the finite-temperature properties of the system. The first quantity we want to address is the superfluid transition temperature $T_c$, which is identified as the lowest temperature at which the superfluid gap $\Delta_{\rm sc}$ vanishes.
Fig.~\ref{delta}(a) shows an example of how the gap varies as the temperature increases. As temperature increases, the superfluid gap $\Delta_{\rm sc}$ decreases monotonically and vanishes at $T_c$.
In contrast, the pseudo-gap $\Delta_{\rm pg}$ is a monotonically increasing function $\Delta_{\rm pg}\sim T^{3/4}$ below $T_c$ and decreases above $T_c$.
On the other hand, Fig.~\ref{delta}(b) shows that the center-of-mass momentum $Q$ of the Cooper pairs is quite insensitive to the temperature for $T<T_c$, and slowly decreases for $T>T_c$. This indicates that the in-plane Zeeman field not only induces a Fulde-Ferrell superfluid below $T_c$, but also induces an exotic normal state above $T_c$ featuring a finite-momentum pseudo-gap \cite{ff4}.

\begin{figure}[htbp]
\centering
\includegraphics[width=0.36\textwidth]{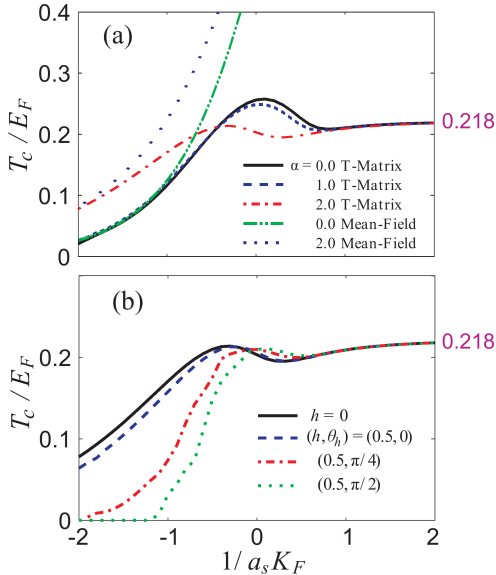}
\caption{(Color online) Critical temperature $T_c$ in the BEC-BCS crossover for (a) different spin-orbit couplings (the values of the SO-coupling strength $\alpha$, in units of $E_F/K_F$, are shown in the legends) with no Zeeman field;
and for (b) different ($h$, $\theta_h$) with $\alpha K_F=2.0E_F$. $h$ is in units of $E_F$.}
\label{crossover}
\end{figure}

In Fig. \ref{crossover}, we show the superfluid transition temperature $T_c$ in the BEC-BCS crossover. In Fig. \ref{crossover}(a), we plot $T_c$ as a function of the interaction strength for several different values of the SO-coupling strength without the Zeeman field. For comparison, the mean-field result is also included. For weak interaction, where $1/(a_sK_F)$ is large and negative, the mean-field result agrees with the T-matrix result. As the interaction increases towards the BEC limit, the mean-field theory predicts an unphysically large critical temperature, a clear indication of its breakdown. In the BEC limit where the two-body attractive interaction is dominant, the system behaves like a condensation of weakly interacting tightly bound bosonic dimers with effective mass $2m$, regardless of the SO coupling strength \cite{jiang_pra2011,zengqiang_prl2011,hu_prl2011}. The BEC transition temperature for this system is 0.218$E_F$ \cite{melo_prl1993,ohashi_prl2002,machida_pra2006}.

\begin{figure}[htbp]
\centering
\includegraphics[width=0.35\textwidth]{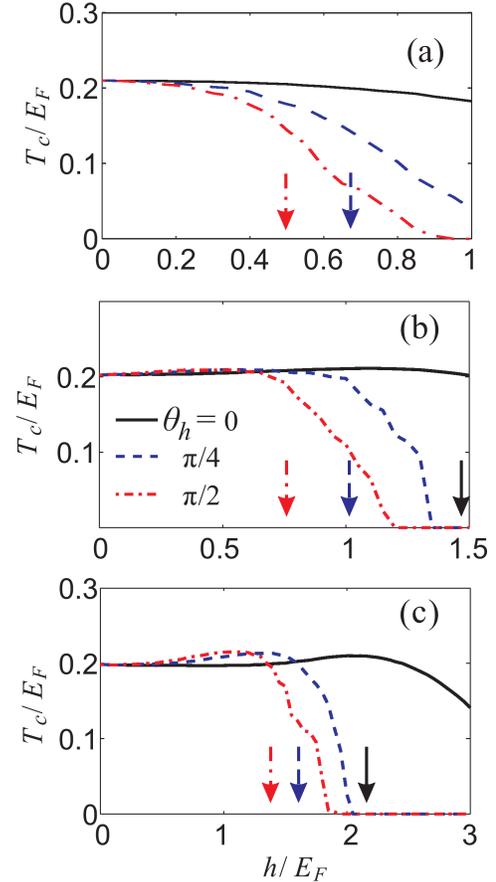}
\caption{(Color online) Critical temperature $T_c$ as functions of the Zeeman field strength for (a) $1/a_sK_F= -0.5$, (b) 0, and (c) 0.5. The solid, dashed, and dot-dashed lines correspond to $\theta_h= 0$, $\pi/4$, and $\pi/2$, respectively. The SO-coupling strength is
$\alpha K_F=2.0E_F$. The vertical arrows indicate the value of the Zeeman field strength at which the system become gapless.}
\label{htc}
\end{figure}

In Fig. \ref{crossover}(b), we examine how the Zeeman field affects $T_c$. In the BEC limit, we have again $T_c=0.218 E_F$, insensitive to either the SO-coupling strength or the Zeeman field strength \cite{jiang_pra2011}. On the BCS side, however, the Zeeman field tends to decrease $T_c$. This effect is much more pronounced with the in-plane Zeeman field than with the out-of-plane Zeeman field, indicating that the finite-momentum Fulde-Ferrell pairing is less robust than the zero-momentum Cooper pairing.

To show the effect of the Zeeman field on the transition temperature more clearly, we plot in Fig.~\ref{htc} $T_c$ as a function of the Zeeman field strength at different orientations. Across the BEC-BCS crossover, $T_c$ is not very sensitive to the out-of-plane Zeeman field over a large range of Zeeman field strengths. By contrast, when the in-plane Zeeman field is present, $T_c$ starts to drop rather sharply near the critical Zeeman field strength where the system becomes gapless. This steep down turn of $T_c$ is particularly pronounced on the BEC side of the resonance. We attribute this steep drop of $T_c$ and the corresponding steep drop of the zero-temperature superfluid gap (see Fig.~\ref{deltah}) to the enhanced fluctuation as a result of the emergence of the nodal surface induced by a large in-plane Zeeman field.

\section{Isothermal Compressibility}
\label{sec_isothermal_compressibility}

The isothermal compressibility, defined as $\kappa = \frac{1}{n} \left. \frac{\partial n}{\partial P} \right|_{T}$, measures the change in the density $n$ in response to the change of the pressure $P$. A recent experiment measured the compressibility of a Fermi gas across the superfluid phase transition \cite{ku_sci2012}, which is found to be in reasonable agreement with the T-matrix theory \cite{haoguo_arxiv2013}. In the superfluid regime, it is found that $\kappa$ increases with temperature, i.e., the gas becomes more compressible as temperature increases. This can be intuitively understood as a lower temperature yields a larger superfluid gap, hence a gas less likely to be compressed. Here we want to examine how SO coupling affects the behavior of $\kappa$.

\begin{figure}[htbp]
\centering
\includegraphics[width=0.48\textwidth]{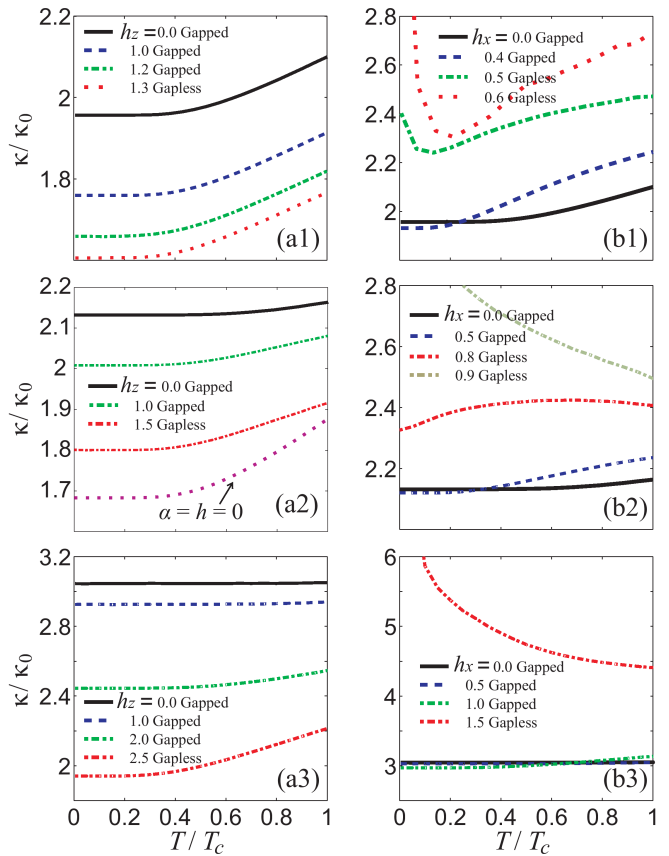}
\caption{(Color online) Compressibility in the superfluid regime as a function of temperature for different Zeeman field strengths and scattering lengths. The scattering lengths are: $1/K_Fa_s = -0.5$ for upper panels (a1) and (b1), $1/K_Fa_s = 0.$ for middle panels (a2) and (b2), and $1/K_Fa_s = 0.5$ for lower panels (a3) and (b3). For left panels (a1), (a2) and (a3), $h_x=0$; for right panels (b1), (b2) and (b3), $h_z=0$. The SO-coupling strength is
$\alpha K_F=2.0E_F$. The compressibility is normalized by $\kappa_0=\frac{3}{2}\frac{1}{N E_F}$, the compressibility of an ideal Fermi gas, and the temperature is normalized to the superfluid transition temperature $T_c$ for the given set of parameters. In the legends, we also indicate whether the quasiparticle excitations of the corresponding system are gapped or gapless. In (a2), we show $\kappa$ in the absence of SO couplings and Zeeman fields (the dotted line) by taking $\kappa=h=0$. In this limit, our calculation recovers the result reported in Ref. \cite{haoguo_arxiv2013}.}
\label{compressibility}
\end{figure}

For our system, $\kappa$ can be expressed as \cite{seo_arxiv2011, seo_pra2013, haoguo_arxiv2013}
\begin{eqnarray}
\kappa &=& \dfrac{1}{N^2}\Big[
\Big(\dfrac{\partial N}{\partial \mu}\Big)_{T,\Delta,Q}
+ \Big(\dfrac{\partial N}{\partial \Delta}\Big)_{T,\mu,Q}  \Big(\dfrac{\partial \Delta}{\partial \mu}\Big)_{T,Q} \notag\\
&& + \Big(\dfrac{\partial N}{\partial Q}\Big)_{T,\mu,\Delta}  \Big(\dfrac{\partial Q}{\partial \mu}\Big)_{T,\Delta}
\Big] ~,
\end{eqnarray}
where
\[
\Big(\dfrac{\partial \Delta}{\partial \mu}\Big)_{T,Q} =
\Big(\dfrac{\partial N}{\partial \Delta}\Big)_{T,\mu,Q} \Big/ \Big(\dfrac{\partial^2 \Omega}{\partial \Delta^2}\Big)_{T,\mu,Q} ~,
\]
\[
\Big(\dfrac{\partial Q}{\partial \mu}\Big)_{T,\Delta} =
\Big(\dfrac{\partial N}{\partial Q}\Big)_{T,\mu,\Delta} \Big/ \Big(\dfrac{\partial^2 \Omega}{\partial Q^2}\Big)_{T,\mu,\Delta} ~.
\]
In Fig. \ref{compressibility}, we show the compressibility $\kappa$ as a function of temperature $T$ in the superfluid regime on both sides of the Feshbach resonance. In the left panels, we set the in-plane Zeeman field $h_x=0$. Across the Feshbach resonance, we see that regardless of the strength of the out-of-plane Zeeman field $h_z$, $\kappa$ increases smoothly as the temperature increases from 0 to $T_c$, just like in a superfluid Fermi gas without SO coupling \cite{ku_sci2012, haoguo_arxiv2013}. By contrast, the presence of the in-plane Zeeman field gives rise to some surprising effects. In the right panels of Fig.~\ref{compressibility}, we set $h_z=0$. One can see that for small $h_x$ such that the system is gapped, $\kappa$ is a monotonically increasing function of $T$. However, once $h_x$ exceeds the critical value such that the system becomes gapless, $\kappa$ becomes a non-monotonic function of temperature. In certain regimes, $\kappa$ even decreases as $T$ increases.

\begin{figure}[htbp]
\centering
\includegraphics[width=0.32\textwidth]{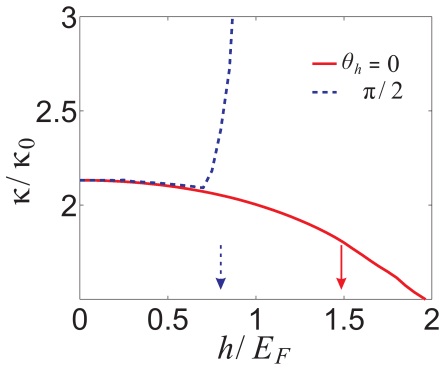}
\caption{(Color online) Compressibility at $T=0$ as a function of the Zeeman field strength. The dashed line corresponds to the in-plane Zeeman field ($\theta_h=\pi/2$), and the solid line corresponds to the out-of-plane Zeeman field ($\theta_h=0$). The dashed and solid arrows indicate the critical field strength at which the system becomes gapless in the presence of an in-plane Zeeman field and an out-of-plane Zeeman field, respectively. The SO coupling strength is
$\alpha K_F=2.0E_F$, and the scatter length is set at $1/K_Fa_s=0$.}
\label{comph}
\end{figure}

The drastically different effects on compressibility by the in-plane and the out-of-plane Zeeman field can also be seen in Fig.~\ref{comph}, where we plot $\kappa$ at the unitarity limit at zero temperature as a function of the Zeeman field strength. For the out-of-plane Zeeman field, $\kappa$ decreases smoothly as the field strength increases. In particular, it does not exhibit any distinctive feature when the system changes from gapped to gapless. This is consistent with the fact that the gapped to gapless transition in the presence of a pure out-of-plane Zeeman field represents a topological phase transition which does not leave a trace in thermodynamic quantities. On the other hand, when the Zeeman field is in plane with its strength increasing from zero, $\kappa$ first decreases and then rises sharply near the critical field when the system becomes gapless. Hence the in-plane Zeeman-field-induced gapless superfluid, featuring a Fermi nodal surface, is highly compressible.

The nodal surface resulting from the in-plane Zeeman field is not unique to Rashba SO coupling. For example, in the experimentally realized equal-weight Rashba-Dresselhaus coupling, a large in-plane Zeeman field can also induce a nodal surface \cite{nodal1,nodal2}. We have checked that for that system, the Zeeman-field dependence of the compressibility exhibits very similar behavior.
As it is known, using the standard form of the fluctuation-dissipation
theorem for a balanced system, the isothermal compressibility can be rewritten as
$\kappa= \frac{V}{T}( \frac{\langle\hat{N}^2\rangle-N^2}{N^2} )$ \cite{seo_arxiv2011}, i.e., $\kappa$ is directly proportional to number fluctuation of the system. The increase in $\kappa$ can therefore be interpreted as a consequence of enhanced number fluctuation induced by the nodal surface.

\section{SUMMARY}
\label{summary}

In summary,
we have investigated the effect of the pairing fluctuation
on the thermodynamic properties of a Rashba SO coupled superfluid Fermi gas by adopting a T-matrix scheme. We focus on the effect of the Zeeman field. In particular, the in-plane Zeeman component leads to finite-momentum Cooper pairing and, when its magnitude becomes sufficiently large, it induces a nodal Fermi surface on which the quasi-particle excitation gap vanishes. The presence of the nodal surface has dramatic effects on the superfluid properties: it greatly suppresses the superfluid transition temperature and increases the isothermal compressibility. Both phenomena can be attributed to the enhanced fluctuation due to the presence of the nodal surface. By stark contrast, when only the out-of-plane Zeeman field is present, both $T_c$ and $\kappa$ exhibit smooth behavior when the Zeeman field strength is increased, even when the system becomes gapless at large Zeeman field strength. The key difference here is that a large out-of-plane Zeeman field, unlike its in-plane counterpart, does not give rise to a nodal surface, but only discrete Fermi points along the $k_z$-axile.
These Fermi points are topological defects, and their appearance does not manifest on any thermodynamic quantities. In addition,
we also find an unconventional pseudo-gap state above the superfluid transition temperature, in which the non-condensed pairs possess nonzero center-of-mass momentum.
We attribute it to the anisotropic Fermi surface induced by the in-plane Zeeman field, which is independent of temperature.

\section*{ACKNOWLEDGEMENTS}
Z.Z., X.Z. and G.G. are supported by the National Natural
Science Foundation of China (Grants No. 11074244
and No. 11274295), and the National 973 Fundamental Research Program (2011cba00200).
H.P. acknowledges support from the NSF, and
the Welch Foundation (Grant No. C-1669).

\appendix
\section{General Formalism}

Introducing the Nambu-Gorkov spinor $\Psi = ( \psi_\uparrow, \psi_\downarrow, \psi_\uparrow^\dag, \psi_\downarrow^\dag )^T$,
the Green's function is given by
\begin{equation}
\mathcal{G}(K) =
\left(
\begin{array}{cc}
G(K) & F(K) \\
\widetilde{F}(K) & \widetilde{G}(K)
\end{array}
\right)~,
\label{green_function}
\end{equation}
where $K=(i\omega_n,\bm{k})$ is the four-dimensional vector and $i\omega_n=(2n+1)\pi T$ is is the fermionic Matsubara frequency.
The relation of the four Green's functions is given by
\begin{eqnarray}
\widetilde{G}( i\omega_n, \bm{k} ) &=& -\Big( G( -i\omega_n, -\bm{k} ) \Big)^T ~, \\
\widetilde{F}( i\omega_n, \bm{k} ) &=& \Big( F( -i\omega_n, \bm{k} ) \Big)^\dag ~.
\end{eqnarray}

\begin{figure}[htbp]
\centering
\includegraphics[width=0.45\textwidth]{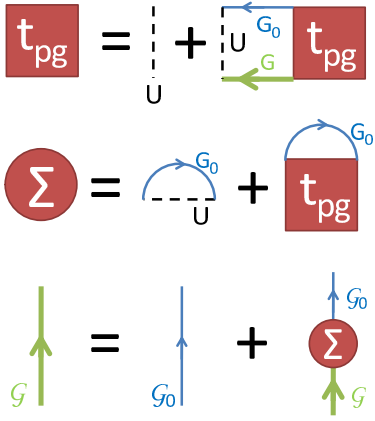}
\caption{(Color online) Feynman diagrams in a T-matrix scheme.}
\label{fig-fd}
\end{figure}

Different from the conventional mean-field theory, the pair propagator consists of two parts in the T-matrix scheme,
\begin{equation}
t(Q') = t_\mathrm{sc}(Q') + t_\mathrm{pg}(Q') ~.
\end{equation}
The superfluid condensate part is
\begin{equation}
t_\mathrm{sc}(Q') = -\Delta_\mathrm{sc}^2\delta(Q'-Q) ~,
\end{equation}
where $Q = ( 0, \bm{Q} )$ and $\bm{Q}$ is the pairing momentum of superfluid. This item comes from the mean-field theory. The pseudo-gap part is introduced by beyond-mean-field effects and in the T-matrix scheme (Fig. \ref{fig-fd}) it is given by
\begin{equation}
t^{-1}_\mathrm{pg}(Q') = U^{-1} + \chi(Q') ~,
\label{t_pg}
\end{equation}
where $\chi(Q')$ is the pair susceptibility in $G_0G$ formalism \cite{lianyi_pra2013},
\begin{equation}
\chi(Q') = \dfrac{1}{2\beta} \mathrm{Tr}\sum\limits_{K} G(K) i\sigma_y \widetilde{G}_0(K-Q') i\sigma_y ~.
\label{chi}
\end{equation}

The full Green's function can be determined as shown in Fig. \ref{fig-fd}.
At $T\leqslant T_c$, the vanishing chemical potential implies \cite{yan_pra2007}
\begin{equation}
t^{-1}_\mathrm{pg}(Q) = U^{-1} + \chi(Q) = 0 ~.
\label{t_pg0}
\end{equation}
It means that $t_\mathrm{pg}(Q')$ behaves like a Dirac $\delta$ function \cite{maly1999, lianyi_pra2013},
\begin{equation}
t_\mathrm{pg}(Q') = -\Delta_\mathrm{pg}^2 \delta(Q'-Q)~.
\label{t_pg_delta}
\end{equation}
The self-energy is given by
\begin{eqnarray}
\Sigma (K, Q) &=& \dfrac{1}{\beta} \sum\limits_{Q'} t(Q') i\sigma_y \widetilde{G}_0(K-Q') i\sigma_y \notag\\
&=& - \Big( \Delta_\mathrm{sc}^2 + \Delta_\mathrm{pg}^2 \Big) i\sigma_y \widetilde{G}_0(K-Q) i\sigma_y \notag\\
&=& - \Delta^2\ i\sigma_y \widetilde{G}_0(K-Q) i\sigma_y ~.
\end{eqnarray}
According to Dyson's equation in the Nambu-Gorkov spinor
$\Psi = \Big( \psi_\uparrow(Q/2+K), \psi_\downarrow(Q/2+K), \psi_\uparrow^\dag(Q/2-K), \psi_\downarrow^\dag(Q/2-K) \Big)^T$,
the full Green's function is
\begin{equation}
\mathcal{G}^{-1}(K, Q) = \mathcal{G}^{-1}_0(K, Q) - \Sigma (K, Q) ~,
\end{equation}

Thermodynamic quantities $\Delta$, $\mu$ and $\bm{Q}$ are obtained by self-consistently solving the Thouless criterion,
\begin{equation}
U^{-1} + \chi(0,\bm{Q}) =0 ~,
\label{thouless}
\end{equation}
the number density conversation,
\begin{equation}
n = \sum\limits_{k\sigma} \psi_\sigma^\dag(k) \psi_\sigma(k) ~,
\label{num-eq}
\end{equation}
and the saddle point of paring susceptibility $\chi(Q)$ \cite{ohashi_jpsj2006},
\begin{equation}
\dfrac{\partial}{\partial \bm{Q}'} \chi(0, \bm{Q}')\Big|_{\bm{Q}'=\bm{Q}} =0 ~.
\label{q-eq}
\end{equation}

The pseudo-gap order parameter can be given as follows.
From Eqs. (\ref{t_pg}) and (\ref{t_pg0}), we can expand $\chi(Q')$ at $Q=(0,\bm{Q})$,
\begin{equation}
\chi(Q') = \chi(Q) + Z\Big( Q_0' - \sum\limits_{i=1}^{3}\dfrac{(\bm{Q}_i' - \bm{Q}_i)^2}{2m_i} \Big) ~,
\end{equation}
where $Q_0' = i\omega_n = 2n\pi T$ is the bosonic Matsubara frequency,
\begin{equation}
Z = \dfrac{\partial \chi(Q')}{\partial Q_0'}\Big|_{Q'=Q} \quad\text{and}
\quad \dfrac{Z}{2m_i} = -\dfrac{1}{2}\dfrac{\partial^2 \chi(Q')}{\partial \bm{Q}_i'^2}\Big|_{Q'=Q} ~.
\end{equation}
Then $t_{pg}(Q')$ is given by
\begin{equation}
t_{pg}(Q')^{-1} = Z\Big( Q_0' - \sum\limits_{i=1}^{3}\dfrac{(\bm{Q}_i' - \bm{Q}_i)^2}{2m_i} \Big) ~.
\label{t_pg_expand}
\end{equation}
Substituting Eq. (\ref{t_pg_expand}) into Eq. (\ref{t_pg_delta}), we obtain
\begin{eqnarray}
\Delta_\mathrm{pg}^2 &=& \dfrac{1}{\beta}\sum\limits_{Q'} \Delta_\mathrm{pg}^2 \delta(Q'-Q)
= - \dfrac{1}{\beta}\sum\limits_{Q'} t_\mathrm{pg}(Q') \notag\\
&=& \dfrac{1}{Z}\prod\limits_{i=1}^{3} \sqrt{\dfrac{T m_i}{2\pi}}\zeta\Big(\dfrac{3}{2}\Big) ~,
\label{delta_pg}
\end{eqnarray}
where $\zeta(3/2)=2.61237535\cdots$.


\begin{thebibliography}{99}

\bibitem{giorgini_rmp2008} S. Giorgini, L. P. Pitaevskii, and S. Stringari, Rev. Mod. Phys. 80, 1215 (2008).

\bibitem{chin_rmp2010} C. Chin, R. Grimm, P. Julienne, and E. Tiesinga, Rev. Mod. Phys. 82, 1225 (2010).

\bibitem{regal_prl2004} C. A. Regal, M. Greiner, and D. S. Jin, Phys. Rev. Lett. \textbf{92}, 040403 (2004).

\bibitem{zwierlein_prl2004} M. W. Zwierlein, C. A. Stan, C. H. Schunck, S. M. F. Raupach, A. J. Kerman, and W. Ketterle,
Phys. Rev. Lett. \textbf{92}, 120403 (2004).

\bibitem{bartenstein_prl2004} M. Bartenstein, A. Altmeyer, S. Riedl, S. Jochim, C. Chin, J. Hecker Denschlag, and R. Grimm,
Phys. Rev. Lett. \textbf{92}, 120401 (2004).

\bibitem{lin_nat2009} Y.-J. Lin, R. L. Compton, K. Jim\'{e}nez-Garc\'{\i}a, J. V. Porto and I. B. Spielman,
Nature \textbf{462}, 628 (2009).

\bibitem{lin_nat2011} Y.-J. Lin, K. Jim\'{e}nez-Garc\'{\i}a and I. B. Spielman,
Nature \textbf{471}, 83 (2011).

\bibitem{ji_nphys2014} Si-Cong Ji, Jin-Yi Zhang, Long Zhang, Zhi-Dong Du, Wei Zheng, You-Jin Deng, Hui Zhai, Shuai Chen and Jian-Wei Pan,
Nature Physics \textbf{10}, 314 (2014).

\bibitem{chunlei_pra2013} Chunlei Qu, Chris Hamner, Ming Gong, Chuanwei Zhang, Peter Engels, Phys. Rev. A, \textbf{88}, 021604 (2013).

\bibitem{pengjun_prl2012} Pengjun Wang, Zeng-Qiang Yu, Zhengkun Fu, Jiao Miao, Lianghui Huang, Shijie Chai, Hui Zhai, and Jing Zhang
Phys. Rev. Lett. \textbf{109}, 095301 (2012).

\bibitem{cheuk_prl2012} Lawrence W. Cheuk, Ariel T. Sommer, Zoran Hadzibabic, Tarik Yefsah, Waseem S. Bakr, and Martin W. Zwierlein
Phys. Rev. Lett. \textbf{109}, 095302 (2012).

\bibitem{gong_prl2011} Ming Gong, Sumanta Tewari, and Chuanwei Zhang, Phys. Rev. Lett. \textbf{107}, 195303 (2011).

\bibitem{han_pra2012} Li Han and C. A. R. S\'{a} de Melo, Phys. Rev. A \textbf{85}, 011606 (2012).

\bibitem{seo_pra2012} Kangjun Seo, Li Han, and C. A. R. S\'{a} de Melo, Phys. Rev. A \textbf{85}, 033601 (2012).

\bibitem{jiang_pra2011} Lei Jiang, Xia-Ji Liu, Hui Hu, and Han Pu, Phys. Rev. A \textbf{84}, 063618 (2011).

\bibitem{zengqiang_prl2011} Zeng-Qiang Yu and Hui Zhai, Phys. Rev. Lett. \textbf{107}, 195305 (2011).

\bibitem{zhoupra2013}X.-F. Zhou, G.-C. Guo, W. Zhang, and W. Yi, Phys. Rev. A {\bf 87}, 063606 (2013).

\bibitem{lindong_njp2013} Lin Dong, Lei Jiang, and Han Pu, New J. Phys. \textbf{15} 075014 (2013).

\bibitem{hu_njp2013} Hui Hu and Xia-Ji Liu, New J. Phys. \textbf{15}, 093037 (2013).

\bibitem{xiaji_pra2012} Xia-Ji Liu, Lei Jiang, Han Pu, Hui Hu, Phys. Rev. A \textbf{85}, 021603 (2012).

\bibitem{wu_prl2013} Fan Wu, Guang-Can Guo, Wei Zhang, and Wei Yi, Phys. Rev. Lett. \textbf{110}, 110401 (2013).

\bibitem{zheng_pra2013} Zhen Zheng, Ming Gong, Xubo Zou, Chuanwei Zhang, and Guangcan Guo, Phys. Rev. A \textbf{87}, 031602 (2013).

\bibitem{chunlei_natc2013} Chunlei Qu, Zhen Zheng, Ming Gong, Yong Xu, Li Mao, Xubo Zou, Guangcan Guo and Chuanwei Zhang,
Nat. Commun. \textbf{4}, 2710 (2013).

\bibitem{weizhang_natc2013} Wei Zhang and Wei Yi, Nat. Commun. \textbf{4}, 2711 (2013).

\bibitem{ff1}L. Dong, L. Jiang, H. Hu, and H. Pu, Phys. Rev. A {\bf 87}, 043616 (2013).

\bibitem{ff4}V. B. Shenoy, Phys. Rev. A {\bf 88}, 033609 (2013).

\bibitem{nozieres1985} P. Nozi\`{e}r\`{e}s and S. Schmitt-Rink, J. Low Temp. Phys. \textbf{59}, 195 (1985).

\bibitem{melo_prl1993} C. A. R. S\'{a} de Melo, Mohit Randeria, and Jan R. Engelbrecht, Phys. Rev. Lett. \textbf{71}, 3202 (1993).

\bibitem{ohashi_prl2002} Y. Ohashi and A. Griffin, Phys. Rev. Lett. \textbf{89}, 130402 (2002)

\bibitem{machida_pra2006} M. Machida and T. Koyama, Phys. Rev. A \textbf{74}, 033603 (2006)

\bibitem{xiaji_pra2005}Xia-Ji Liu and Hui Hu, Phys. Rev. A \textbf{72}, 063613 (2005).

\bibitem{qijin_prl1998} Qijin Chen, Ioan Kosztin, Boldizs\'{a}r Jank\'{o}, and K. Levin, Phys. Rev. Lett. \textbf{81}, 4708 (1998).

\bibitem{maly1999} Jiri Maly, Boldizs\'{a}r Jank\'{o}, K. Levin, Physica C \textbf{321}, 113 (1999).

\bibitem{loktev_pr2001} V. M. Loktev, R. M. Quick, and S. G. Sharapov, Phys. Rep. \textbf{349}, 1 (2001).

\bibitem{perali_prb2007} A. Perali, P. Pieri, G. C. Strinati, and C. Castellani, Phys. Rev. B \textbf{66}, 024510 (2002).

\bibitem{qijin_pr2005} Q. J. Chen, J. Stajic, S. N. Tan, and K. Levin, Phys. Rep. \textbf{412}, 1 (2005).

\bibitem{kinnunen} J. Kinnunen, M. Rodriguez, and P. T\"{o}rm\"{a}, Science {\bf 305}, 1131 (2004).

\bibitem{huihu_prl2010} H. Hu, X.-J. Liu, P. D. Drummond, and H. Dong, Phys. Rev. Lett. \textbf{104}, 240407 (2010).

\bibitem{qijin}Q. Chen, Y. He, C.-C. Chien, and K. Levin, Rep. Prog. Phys. {\bf 72},
122501 (2009).

\bibitem{kinast}J. Kinast, A. Turlapov, J. E. Thomas, Q. J. Chen, J. Stajic, and K. Levin,  Science {\bf 307}, 1296 (2005).

\bibitem{stajic}J. Stajic, Q. J. Chen, and K. Levin, Phys. Rev. Lett. {\bf 94}, 060401 (2005).



\bibitem{bauer_prl2014} Marianne Bauer, Meera M. Parish, and Tilman Enss, Phys. Rev. Lett. \textbf{112}, 135302 (2014).

\bibitem{haoguo_arxiv2013} Hao Guo, Yan He, Chih-Chun Chien, K. Levin, Phys. Rev. A {\bf 88}, 043644 (2013).

\bibitem{lianyi_pra2013} Lianyi He, Xu-Guang Huang, Hui Hu, and Xia-Ji Liu, Phys. Rev. A \textbf{87}, 053616 (2013).

\bibitem{zhangjing}Zhengkun Fu, Lianghui Huang, Zengming Meng, Pengjun Wang, Xia-Ji Liu, Han Pu, Hui Hu, and Jing Zhang, Phys. Rev. A {\bf 87}, 053619 (2013).

\bibitem{renyuan_prl2012} Renyuan Liao, Yu Yi-Xiang, and Wu-Ming Liu, Phys. Rev. Lett. \textbf{108}, 080406 (2012).

\bibitem{lianyi_prl2012} Lianyi He and Xu-Guang Huang, Phys. Rev. Lett. \textbf{108}, 145302 (2012).

\bibitem{tewari_prl2007} Sumanta Tewari, S. Das Sarma, Chetan Nayak, Chuanwei Zhang, and P. Zoller,
Phys. Rev. Lett. \textbf{98}, 010506 (2007).

\bibitem{chuanwei_prl2012} Chuanwei Zhang, Sumanta Tewari, Roman M. Lutchyn, and S. Das Sarma, Phys. Rev. Lett. \textbf{101}, 160401 (2008).

\bibitem{seo_pra2013} Kangjun Seo, Chuanwei Zhang, and Sumanta Tewari, Phys. Rev. A \textbf{87}, 063618 (2013).

\bibitem{kosztin_prb2000} Ioan Kosztin, Qijin Chen, Ying-Jer Kao, and K. Levin, Phys. Rev. B \textbf{61}, 11662 (2000).

\bibitem{thouless_1960} D. J. Thouless, Ann. Phys. \textbf{10}, 553 (1960).

\bibitem{hu_prl2011} Hui Hu, Lei Jiang, Xia-Ji Liu, Han Pu, Phys. Rev. Lett. \textbf{107}, 195304 (2011).

\bibitem{hu_pra2012} Hui Hu, Han Pu, Jing Zhang, Shi-Guo Peng, Xia-Ji Liu, Phys. Rev. A \textbf{86}, 053627 (2012).

\bibitem{yan_pra2007} Yan He, Chih-chun Chien, Qijin Chen, and K. Levin, Phys. Rev. A \textbf{75}, 021602 (2007).

\bibitem{yong_prl2014} Yong Xu, Rui-Lin Chu, Chuanwei Zhang, Phys. Rev. Lett. \textbf{112}, 136402 (2014).

\bibitem{ohashi_jpsj2006} Yoji Ohashi,	J. Phys. Soc. Jpn. \textbf{71} 2625 (2002).

\bibitem{ku_sci2012}M. J. H. Ku, A. T. Sommer, L. W. Cheuk, and M. Zwierlein, Science {\bf 335}, 563 (2012).



\bibitem{seo_arxiv2011} Kangjun Seo, Carlos A. R. S\'{a} de Melo, arXiv: 1105.4365 (2011).

\bibitem{nodal1}X.-J. Liu, and H. Hu, Phys. Rev. A {\bf 87}, 051608 (2013).

\bibitem{nodal2}Fan Wu, Guang-Can Guo, Wei Zhang, and Wei Yi, Phys. Rev. A {\bf 88}, 043614 (2013).

\end{thebibliography}
\end{document}